\documentclass[a4paper, 12pt, oneside]{article}

\pdfoutput=1

\usepackage[utf8]{inputenc}
\usepackage[english]{babel}
\usepackage{graphicx}
\usepackage[margin=1in]{geometry}
\usepackage{rotating}
\usepackage[numbers]{natbib}
\usepackage{caption,tabularx}
\usepackage{latexsym}
\usepackage{slashbox}
\usepackage{multirow}
\usepackage{authblk}
\usepackage{bbm}
\usepackage{comment}
\usepackage{dcolumn}
\usepackage{url}
\usepackage{hyperref}
\usepackage{booktabs}
\usepackage{array}
\usepackage{adjustbox}
\usepackage[usenames,dvipsnames,svgnames,table]{xcolor}
\usepackage[toc,page]{appendix}
\usepackage[linesnumbered,ruled]{algorithm2e}
\usepackage{tikz}
\usetikzlibrary{shapes.geometric, arrows, calc, fit, positioning, chains, arrows.meta}

\usepackage{subcaption}
\usepackage[linesnumbered,ruled]{algorithm2e}

\usepackage{amsmath, amsthm, amsfonts}
\usepackage{amssymb}

\title {The hidden waves in the ECG uncovered:\\
a  sound automated interpretation method
 }

\author[1,*]{Cristina Rueda}
\author[1]{Yolanda Larriba}
\author[1]{Adri\'an Lamela}
\date{}

\affil[1]{Department of Statistics and Operations Research,
Universidad de Valladolid, Valladolid, Spain}

\theoremstyle{definition}
   
   \newtheorem{Def}{Definition}

\begin{document}
\maketitle

\begin{abstract}
 A novel approach for analysing cardiac rhythm data is presented in this paper. Heartbeats are decomposed into the five fundamental $P$, $Q$, $R$, $S$ and $T$ waves plus an error term to account for artefacts in the data which provides a meaningful, physical interpretation of the heart's electric system. The morphology of each wave is concisely described using four parameters that allow to all the different patterns in heartbeats  be characterized and thus differentiated

This multi-purpose approach solves such questions as the extraction of interpretable features, the detection of the fiducial marks of the fundamental waves, or the generation of synthetic data and the denoising of signals. Yet, the greatest benefit from this new discovery will be the automatic diagnosis of heart anomalies as well as other clinical uses with great advantages compared to the rigid, vulnerable and black box machine learning procedures, widely used in medical devices.

The paper shows the enormous potential of the method in practice; specifically, the capability to discriminate subjects,  characterize morphologies and detect the fiducial marks (reference points) are validated numerically using simulated and real data, thus proving that it outperforms its competitors.

\end{abstract}

\section{Introduction}

\begin{figure}[h!]
\begin{subfigure}{\linewidth}  
  \centering
  \includegraphics[width=0.9\textwidth]{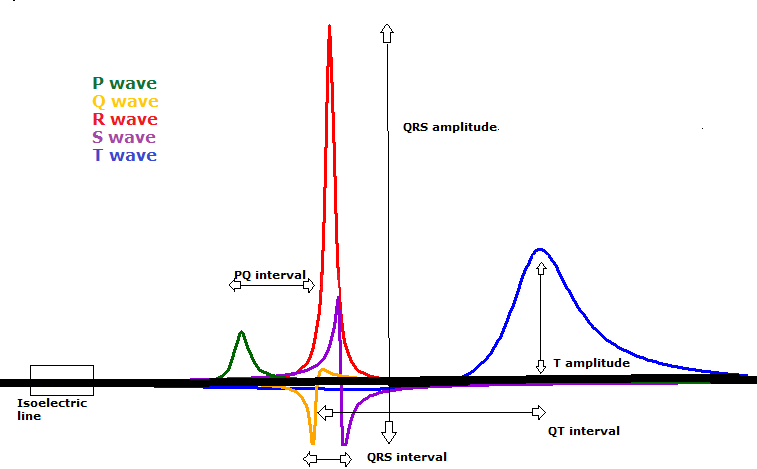}
  \caption{}
  \label{f:Figure0}
\end{subfigure}\\
\begin{subfigure}{\linewidth}  
  \centering
  \includegraphics[width=0.9\textwidth]{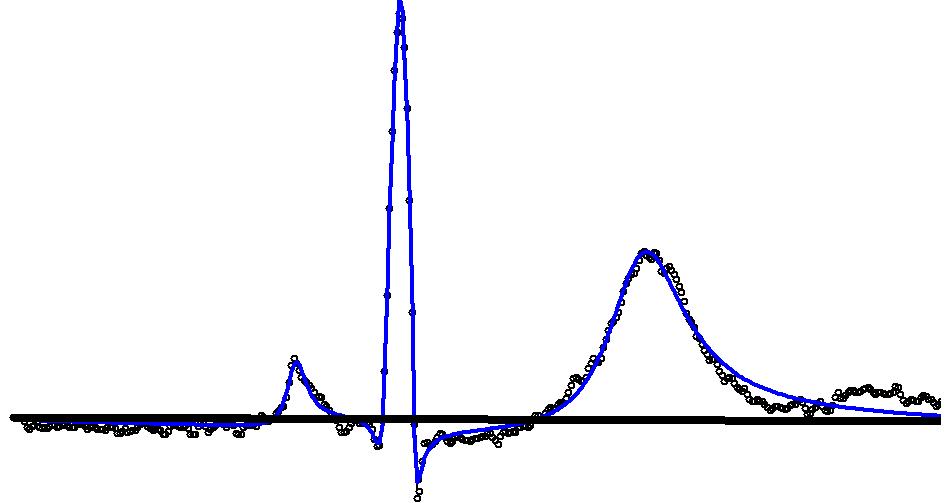}
  \caption{}
  \label{f:Figure1}
\end{subfigure}%
\caption{ (a) The five waves : $P$,$Q$,$R$,$S$,$T$ derived from the $FMM_{ecg}$ model and some of the main features that are derived from the parameters of the model in a simple way. (b) Observed signal (black points) and $FMM_{ecg}$ fit (blue). Data from  patient \textit{sel}106 from MIT-BIT Arrhythmia Database  from Physionet ({http://www.physionet.org}) }
\label{f:Figure0and1}
\end{figure}

The importance of the ECG signal in diagnosis and prediction of cardiovascular diseases is worth noting. 
The process recorded in the ECG is the periodic electrical activity of the heart. This activity represents the contraction and relaxation of the atria and ventricle, processes related to the crests and troughs of the ECG waveform, labelled $P$, $Q$, $R$, $S$ and $T$ (see Figure \ref{f:Figure0and1} (a)). The main features used in the medical practice are related to the location and amplitudes of these waves. A standard ECG signal is registered using twelve leads calculated from different electrodes being Lead II the reference one. 
\\

The mere visual observation of the ECG signals, although made by a consolidated expert, is not enough to discover the diversity of abnormalities and the specific characteristics of the morphology of each ECG. Moreover, it requires an enormous amount of human expertise resources. Therefore, a rigorous automatic analysis of digitalized ECG signals can be of great help. However, although it has been a question that has received a lot of attention in the literature over the last decades, there is still no suitable mathematical model or computational approach, that accurately describes the spectrum of morphologies in  ECG signals, as is noted in recent references on this topic, such as \cite{Mar17}, \cite{Qua17},  \cite{Kir17}, \cite{Sch17}  or  \cite{Din19}, among others.


The literature addressing the problem of the automatic interpretation of the ECG is so extensive that it is difficult to include a complete review here. The most widely used model-based approach describes the main waves with a combination of basic functions, the Gaussians being the preferred ones, for a single or average beat. A precursor model was proposed  by \cite{Mcs03} and was more recently considered by \cite{Say10} or \cite{Roo13} among others, whom proposed improvements in the formulation and estimation algorithms. \cite{Pen19} also recently uses this approach for the predictive modelling of drug effects on ECG signals. These approaches have important shortcomings. In particular, the Gaussian functions fail to reproduce the morphology of the waves in a simple way, especially for atypical and noisy ECGs, where the complexity as well as the risk of over fitting increase. Moreover, most of the parameters do not have a specific morphological meaning.
Other examples of model-based proposals are those by \cite{Lop18}, \cite{Qui19}, \cite{Kot17}, \cite{Sha18} or \cite{Rak18}. These approaches may be suitable to study some specific questions, but, they are far from being multi-purpose methods.

However, many of the recent papers are contributions to  computational and machine learning approaches. Some of the large list of references are: \cite{Roo17}, \cite{Ber18}, \cite{Pla18}, \cite{Fri18}, \cite{Han19},  \cite{Min19} or \cite{Tis19}.  Also, the papers by \cite{Sch17}, \cite{Luz16}, \cite{Lyo18}, \cite{Tei18},  \cite{Att19} and \cite{Sev20} extended the list of procedures and their pros and cons for the automatic analysis of ECGs. 
  In general, machine learning approaches success is very dependent on the training set, the selection of diagnostic groups, the preprocessing and the data base. Furthermore, they are rigid and black-box procedures that are susceptible to adversial attacks \cite{Han20}.
  
 The approach, called $FMM_{ecg}$, presented in this paper is just the opposite.
 \\

This novel approach combines a physically meaningful formulation with good statistical and computational properties. $FMM_{ecg}$ is a multicomponent model, where each component is a single $FMM$ (Frequency Modulate M\"obius) oscillator and specific ECG parameter restrictions are included. Single $FMM$ models are recently proposed by  \cite{Rue19} to predict oscillatory signals in several different fields from biology to astrophysics. The distinguishing feature of the $FMM$ model is that it is formulated in terms of the phase, which is the angular variable that represents the periodic movement of the oscillation.
Specifically, the $FMM_{ecg}$ model is defined as the combination of exactly five oscillatory components referred to as waves: $W_J(),  J = P, Q, R, S, T$, which correspond to the fundamental waves in a heartbeat; plus an error term that accounts for artefacts in the data. Four parameters characterize each wave and, a Maximization-Identification (MI) algorithm is designed to estimate them. This algorithm alternates, iteratively, between a maximization M-step and a wave-identification I-step. While the model proposal is valid for signals registered elsewhere, the I-step is lead-specific. Nevertheless, the I-step can be easily adapted to signals registered in other regions.

The main virtues of the novel approach can be summarized in five points which are validated in the paper. Firstly, the $FMM_{ecg}$ model is physically meaningful representing the conduction of the electrical signal by the combination of five main waves presented in a normal heartbeat. Therefore, alterations in a specific wave identifies the part of the heart responsible. Secondly, for each wave, four parameters are extracted, measuring, amplitude, location, scale and shape. These parameters are able to characterize, reproduce and identify the variety of morphologies observed in real ECG signals. In addition, other interesting features are easily derived from these main parameters. Thirdly, the MI algorithm provides accurate and robust estimates of the model parameters discarding overfitting problems. Fourthly, the approach is not dependent on a training set and is valid for any ECG registered signal, independently of the preprocessing, frequency or scale. Finally, the approach has strong theoretical properties: is maximum likelihood based while assuming Gaussian errors, the parameters are identifiable and the estimators are consistent.
\\

The validation of  the $FMM_{ecg}$ approach is not simple as there are  many properties that the model is supposed to verify. Moreover, there is no multi-purpose approach in the literature similar to  $FMM_{ecg}$. Therefore, the main properties of $FMM_{ecg}$ are validated considering diverse alternative approaches. On the one hand, for global goodness of fit consistency, robustness and discriminative power, the $FMM_{ecg}$ is compared with a model-based approach, which considers a combination of Gaussian components, similar to that proposed by \cite{Roo13}. On the other hand, the ability to detect fiducial marks is compared with several recent machine learning approaches, in particular, those considered by \cite{Fri18}. In this paper, we deal with signals from Lead II and close to it. Simulated and publicly available data from databases in Physionet (www.physionet.org) \cite{Gol00} are used. Very promising results have been obtained from real data. For example, Figure 1 shows the result of applying the $FMM_{ecg}$ to data from patient \textit{sel106} in MIT-BIT database, a representative, typical pattern used by many authors. The waves drawn in Figure 1 (a) have not been artificially generated, but are simply the estimators provided by the MI algorithm for the five waves: $W_J (), J = P, Q, R, S, T$. While panel (b) shows the combined $FMM_{ecg}$ fit.

\section{Overview of the $FMM_{ecg}$ model}

Suppose  $X(t_i), t_1<...<t_n $ are observations from one beat. 
Without loss of generality, we assume that  $ t_i\in [0,2\pi]$ (in any other case, transform the observed time points as in \cite{Rue19}.

For $J \in\lbrace P,Q,R,S,T\rbrace $, let  $\upsilon_J=(A_J,\alpha_J,\beta_J,\omega_J)'$ be the four-dimensional parameters describing the waveforms in such a way that
 \begin{equation*}
   W_J(t,\upsilon_J)=A_J\cos(\beta_J+2\arctan(\omega_J\tan(\frac{t-\alpha_J}{2})))
   \end{equation*}
Then, the $FMM_{ecg}$ model, is defined as a parametric additive signal plus error model as follows:
\begin{Def}{ \textit{$FMM_{ecg}$ model}} \label{eq:Mobecg}.
For $i=1,...,n$:
 \begin{equation*}
 X(t_i)= \mu(t_i,\theta)+ e(t_i); 
\end{equation*}
where,
 \begin{equation*}
\mu(t,\theta)= M+\sum_{J\in \lbrace P,Q,R,S,T \rbrace} W_J(t,\upsilon_J);
\end{equation*}
and
\begin{itemize}
\item  $\theta=(M,\upsilon_P,\upsilon_Q.\upsilon_R,\upsilon_S,\upsilon_T)$ verifying:
\begin{enumerate}
\item $ M \in \Re $
\item $\upsilon_J \in    \Re^+ \times [0,2\pi] \times [0,2\pi] \times[0,1]$, $J \in \lbrace P,Q,R,S,T \rbrace $

\item   $\alpha_P \le \alpha_Q \le \alpha_R \le \alpha_S \le \alpha_T \le \alpha_P $  
\end{enumerate}
\item $(e(t_1),...,e(t_n))' \sim N_n(0,\sigma^2I)$

\end{itemize}
\end{Def}

 The incorporation of circular order restrictions among the $\alpha$'s represent the ordered movement of the stimulus from the sinus node to the ventricles, passing through the atria, this giving the model physical interpretability. The restrictions guarantee the identifiability of the parameters once main wave $R$ is located.
\\

The parameter $M$ is an intercept parameter and the components of $\upsilon_J$ describe different aspects of the morphology of wave $J$. Specifically, the parameter $A_J$ measures the wave amplitude; a zero value indicating that the corresponding wave is not present.  The parameter $\alpha_J$ is a location parameter. In addition, $\beta_J$ and $\omega_J$ measure skewness and kurtosis, respectively. More specifically, assuming $\alpha_J=0$, the values for parameter $\beta_J$ close to $\pi$ (or $2\pi$) represent a unimodal symmetric wave (or an inverse unimodal symmetric wave); as $\beta_J$ moves away from these values, the patterns are more asymmetric and the values of $\beta_J$ equal to $\pi/2$ or $3\pi/2$ describe a wave with both crest and trough with completely asymmetric  patterns. The parameter $\omega_J$ measures the sharpness of the peak, $\omega_J=1$ corresponds to an exact sinusoidal shape and, as $\omega_J$ approaches zero, the sharpness becomes more pronounced (see \cite{Rue19} for more details in parameter interpretation). 
 \\
 
 Other  features extracted from the main parameters are the marks for  crest ($t^U$) and trough times ($t^L$), defined for $J\in \lbrace P,Q,R,S,T \rbrace$ as follows:
\begin{equation*}
t^U_J=\alpha_J+2 \arctan (\frac{1}{\omega_J}\tan(\frac{-\beta_J}{2}));
t^L_J=\alpha_J+2 \arctan (\frac{1}{\omega_J}\tan(\frac{\pi-\beta_J}{2}))
\end{equation*}

Moreover, measurements of inter-wave intervals, as those in Figure \ref{f:Figure0and1} (a) are calculated using angular distances between these marks, and other features, such as those used in the literature of ECG interpretation, can be easily derived from the main parameters. However, while the estimation of features proposed in the literature often depends on the algorithm and voltage measurements  \cite{Sch17}, $FMM_{ecg}$ provides systematic and reliable measurements. 

In the estimation process, to improve the waves identification when atypical patterns are observed, additional conditions are imposed.

The dependence of signal, waves and model on the parameters $\theta$ or $\upsilon$ is omitted when no confusion across this paper. 

\section{Validation}
 Three different validation analyses have been performed.  The first two refer to the QT database \cite{Lag97} and the third is a simulation experiment, which is deferred to Supplementary Information. The QT database was chosen as it has been used recently by several authors and provides a wide range of morphologies associated with healthy and pathological ECG's. The database contains 105 ECG records and signals from two leads. We analyse the segment for each patient for whom the $T$ or $P$ waves have been manually annotated, as well as the data corresponding to the signal closest to Lead II (in most cases it corresponds to the first signal). For patient \textit{sel}42, data from the first signal are not reliable, instead, the inverse of the second one is analysed as it represents a signal closer to Lead II. 

A total of 3,623 single beats signals have been analysed.

The validation includes, the global fit of the model, the identifiability of parameters, the accuracy and consistency of estimators, the robustness of the model against noise, the capability to characterize different morphologies, but also the performance in specific tasks of practical interest as the subject discrimination or the determination of the fiducial marks of $T$ and $P$ waves.

\begin{figure}[htbp]
\begin{subfigure}{0.5\linewidth}  
  \centering
  \includegraphics[width=\textwidth]{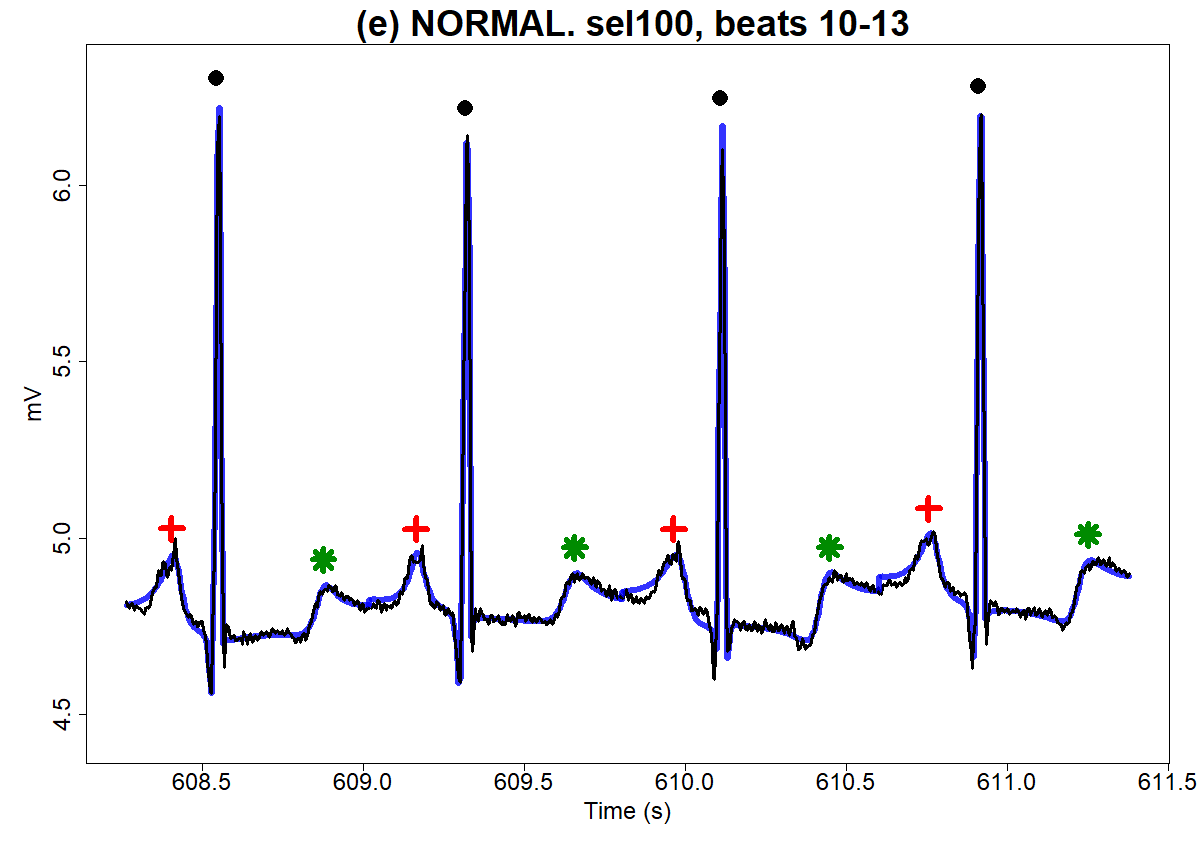}
  \label{f:Figure2a}
\end{subfigure}
\begin{subfigure}{.5\linewidth}  
  \centering
  \includegraphics[width=\textwidth]{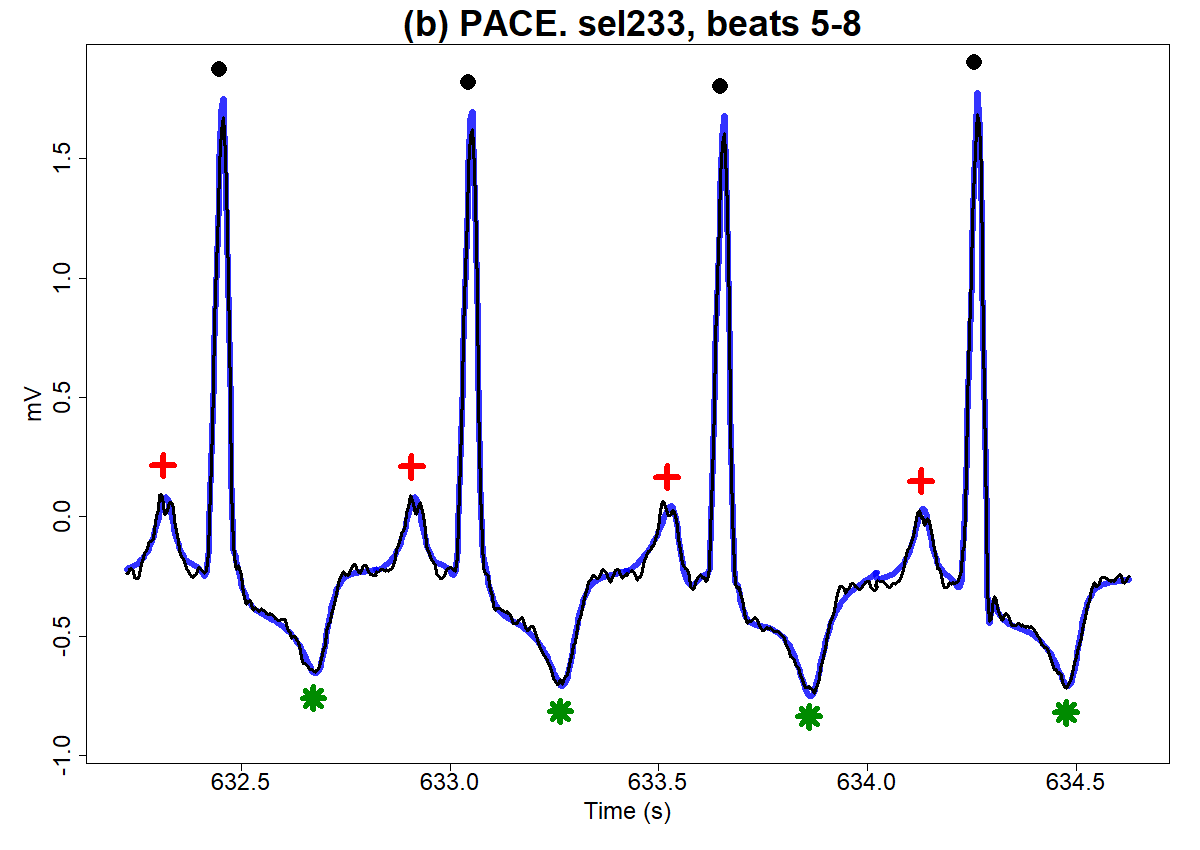}
  \label{f:Figure2b}
\end{subfigure}\\
\begin{subfigure}{.5\linewidth}  
  \centering
  \includegraphics[width=\textwidth]{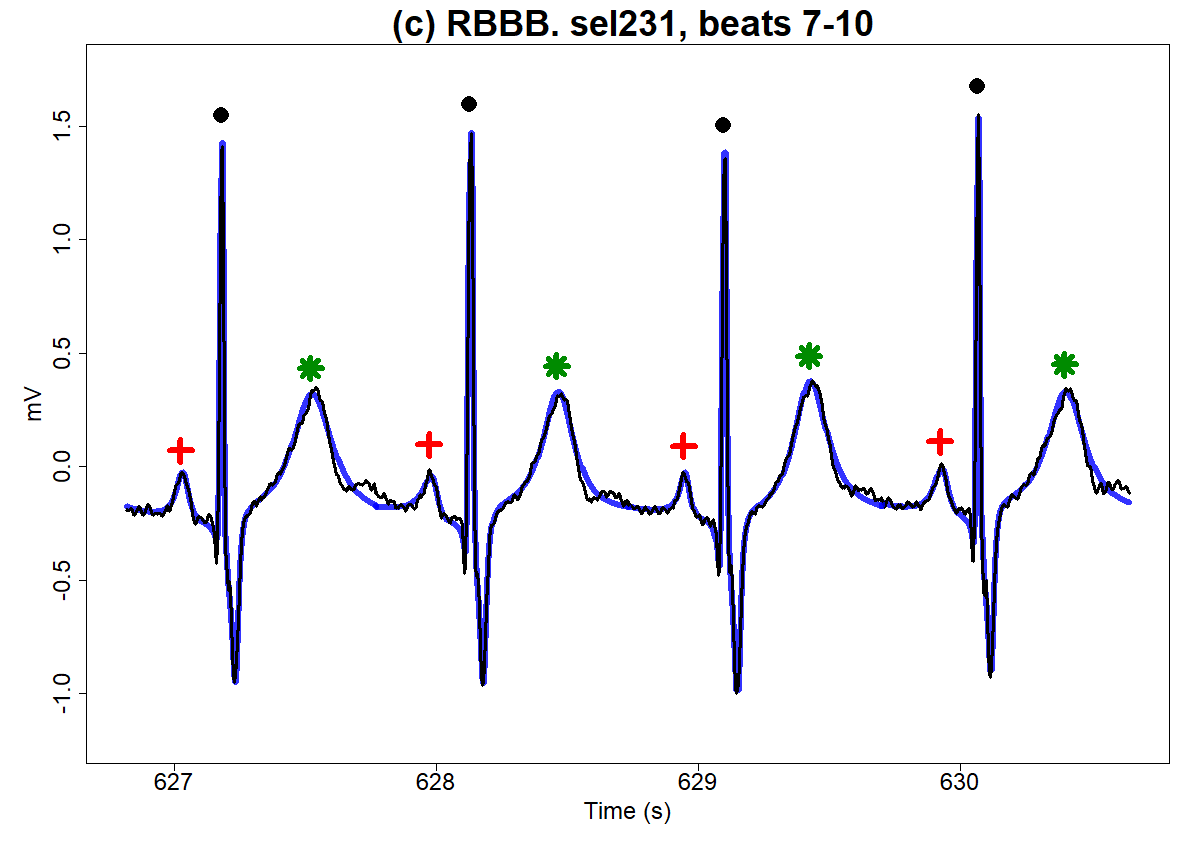}
  \label{f:Figure2c}
\end{subfigure}
\begin{subfigure}{.5\linewidth}  
  \centering
  \includegraphics[width=1\textwidth]{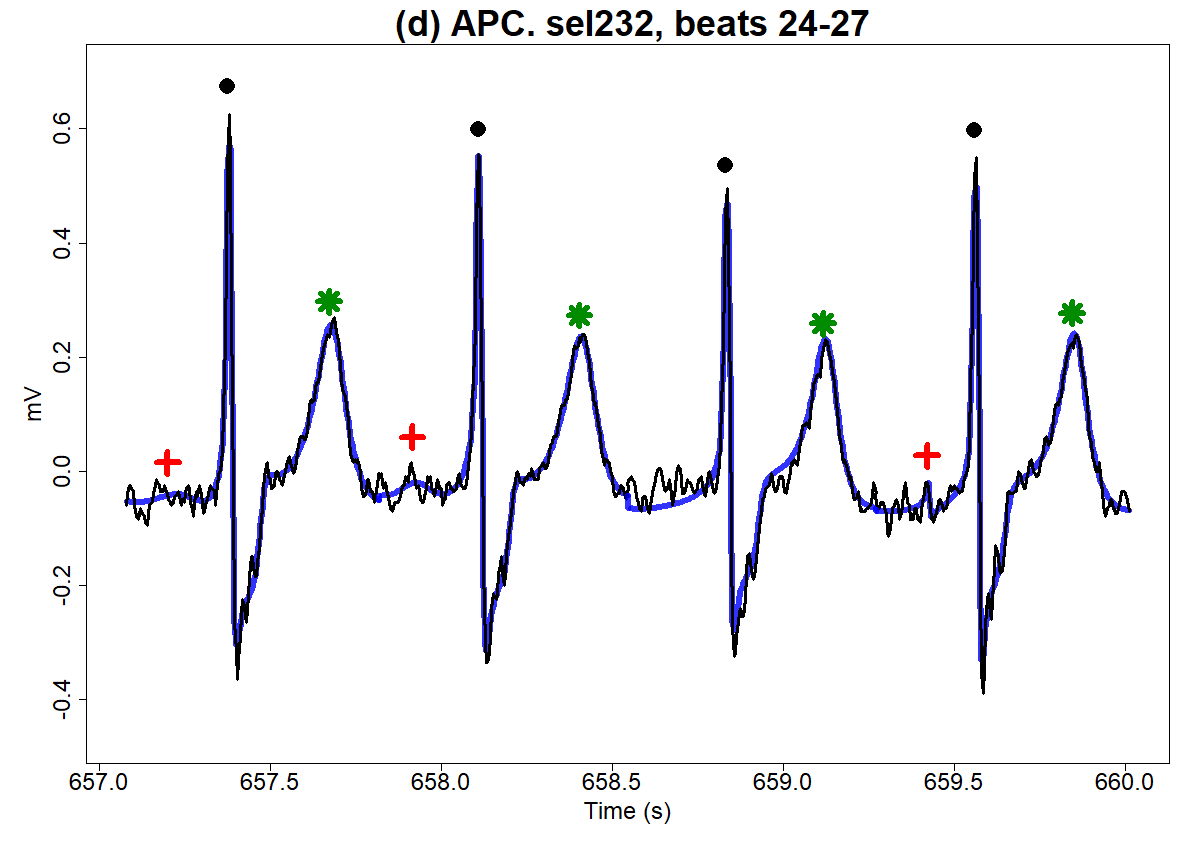}
  \label{f:Figure2d}
\end{subfigure}\\
\begin{subfigure}{.5\linewidth}  
  \centering
  \includegraphics[width=1\textwidth]{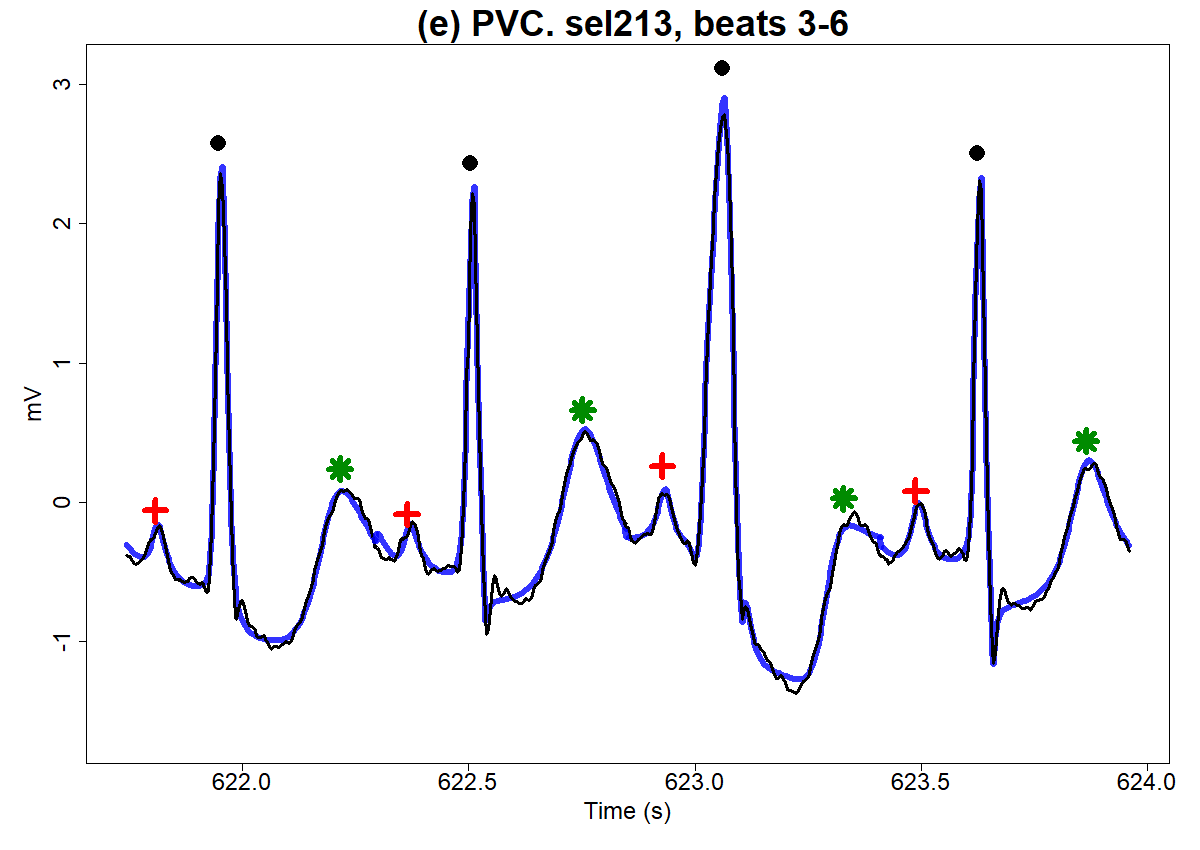}
  \label{f:Figure2e}
\end{subfigure}
\begin{subfigure}{.5\linewidth}  
  \centering
  \includegraphics[width=1\textwidth]{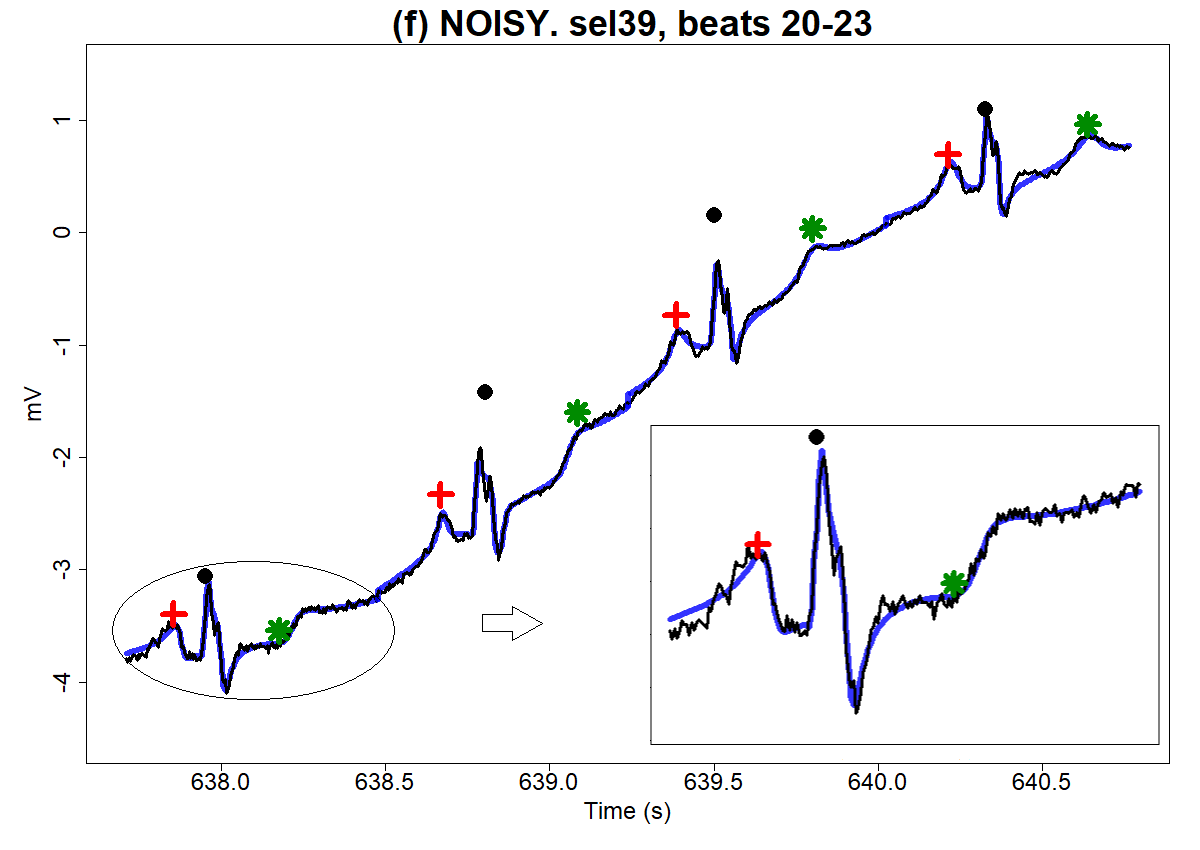}
  \label{f:Figure2f}
\end{subfigure}%
\caption{Observed ECG segments (black lines), $FMM_{ecg}$  fits (blue lines) and fiducial marks for $R$ wave( $\bullet$), $T$ wave({\color{green} $\star$ }), $P$ wave({\color{red}$+$}); for (a) NORMAL, (b) PACE, (c) RBBB, (d) APC, (e) PVC and (f) NOISY patterns.}
\label{f:Figure2}
\end{figure}
\subsection{Analysis of QT database signals. Graphical and analytical results.}
For each single beat, the value of a coefficient of determination that measures the proportion of the variance explained by the model out of the total variance, is denoted by $R^2$ and is obtained as follows:
\begin{equation}\label{eq:R}
R^2=1-\frac{\sum_{i=1}^n (X(t_i)-\hat{\mu}(t_i))^2}{\sum_{i=1}^n (X(t_i)-\overline{X})^2}
\end{equation}

The $R^2$ values are very high across patients, being  $R^2$ global mean (SD) equal to 0.98(0.02). 

A selection of ECG segments to which the model has been fitted, are shown in Figure \ref{f:Figure2}, the first five correspond to the most frequent categories according to Physionet's classification of the heartbeats by their morphology. The selected categories  are the ones that appear most frequently in the databases and are identified as: NORMAL (typical pattern), PACE (Paced beat), RBBB (Right bundle branch block beat), APC (Atrial premature beat), and PVC (Premature ventricular contraction); besides a NOISY pattern is also considered. The NOISY pattern exhibited both, low and high frequency noise as the zoom in the corresponding plot shows. The $R^2$ specific means, is equal to 0.92 for the NOISY and higher than 0.98 for the others. 

\begin{figure}[htbp]
\begin{subfigure}{0.5\linewidth}  
  \centering
  \includegraphics[width=\textwidth]{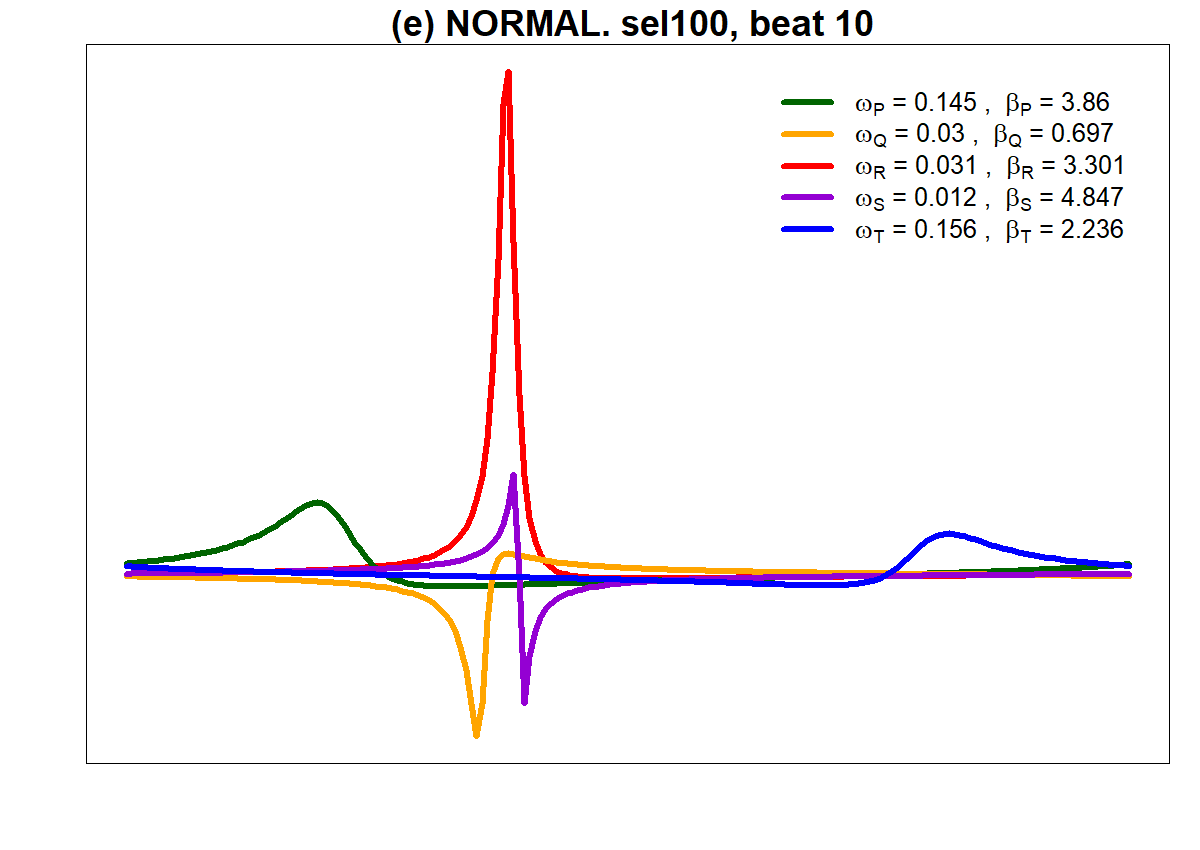}
  \label{f:Figure2sa}
\end{subfigure}
\begin{subfigure}{.5\linewidth}  
  \centering
  \includegraphics[width=\textwidth]{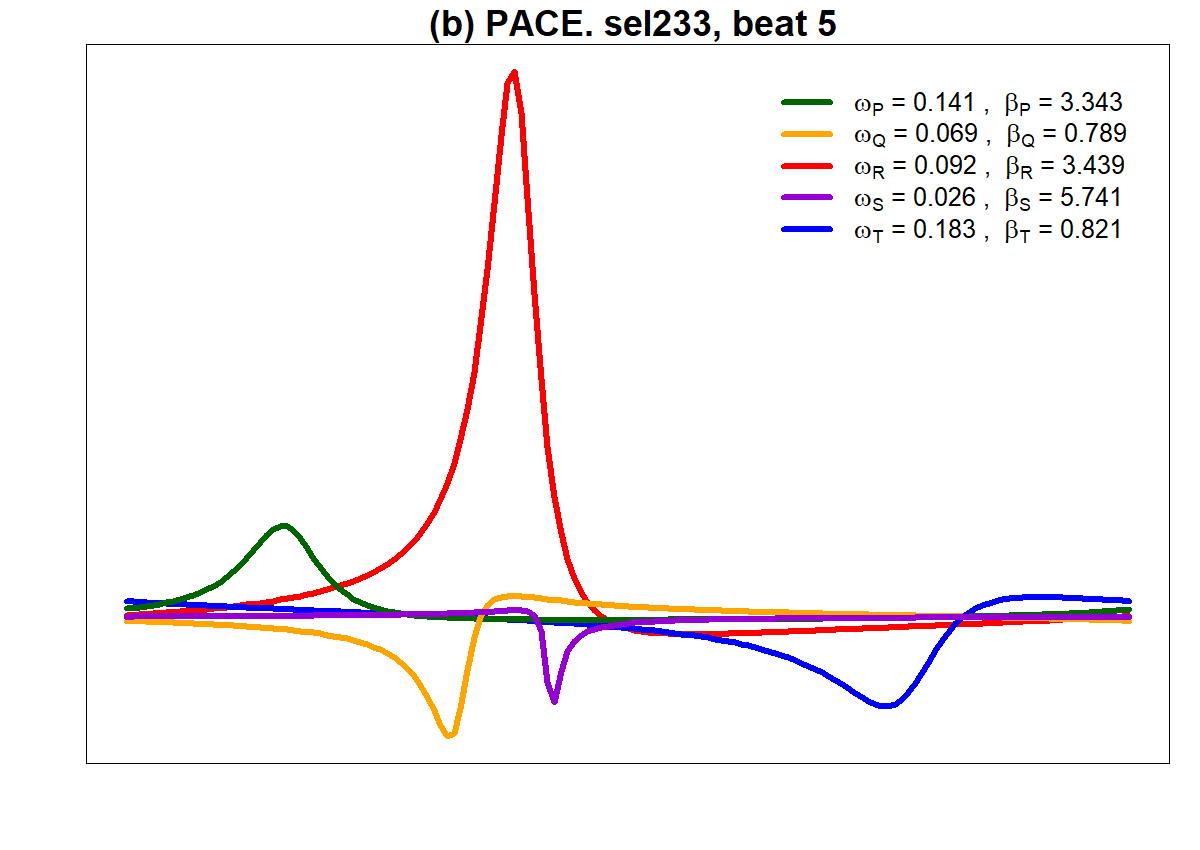}
  \label{f:Figure2sb}
\end{subfigure}\\
\begin{subfigure}{.5\linewidth}  
  \centering
  \includegraphics[width=\textwidth]{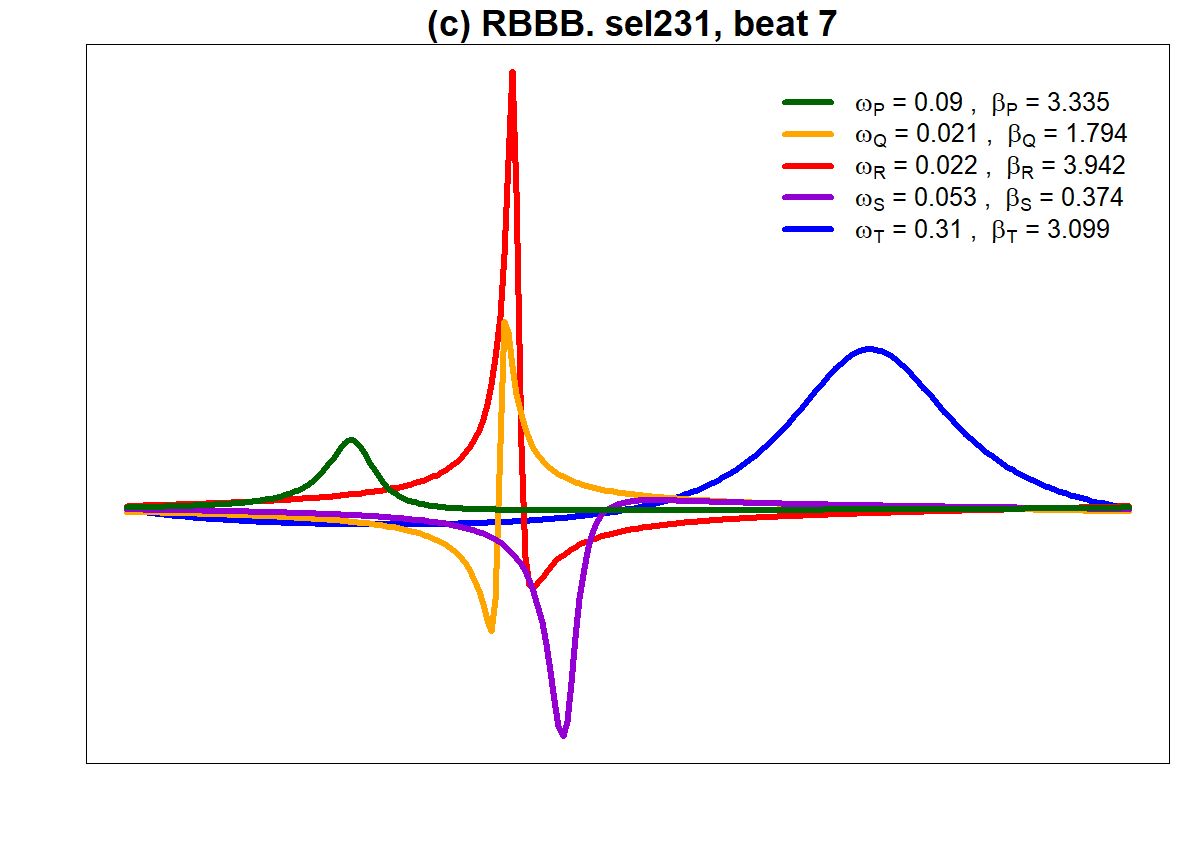}
  \label{f:Figure2sc}
\end{subfigure}
\begin{subfigure}{.5\linewidth}  
  \centering
  \includegraphics[width=1\textwidth]{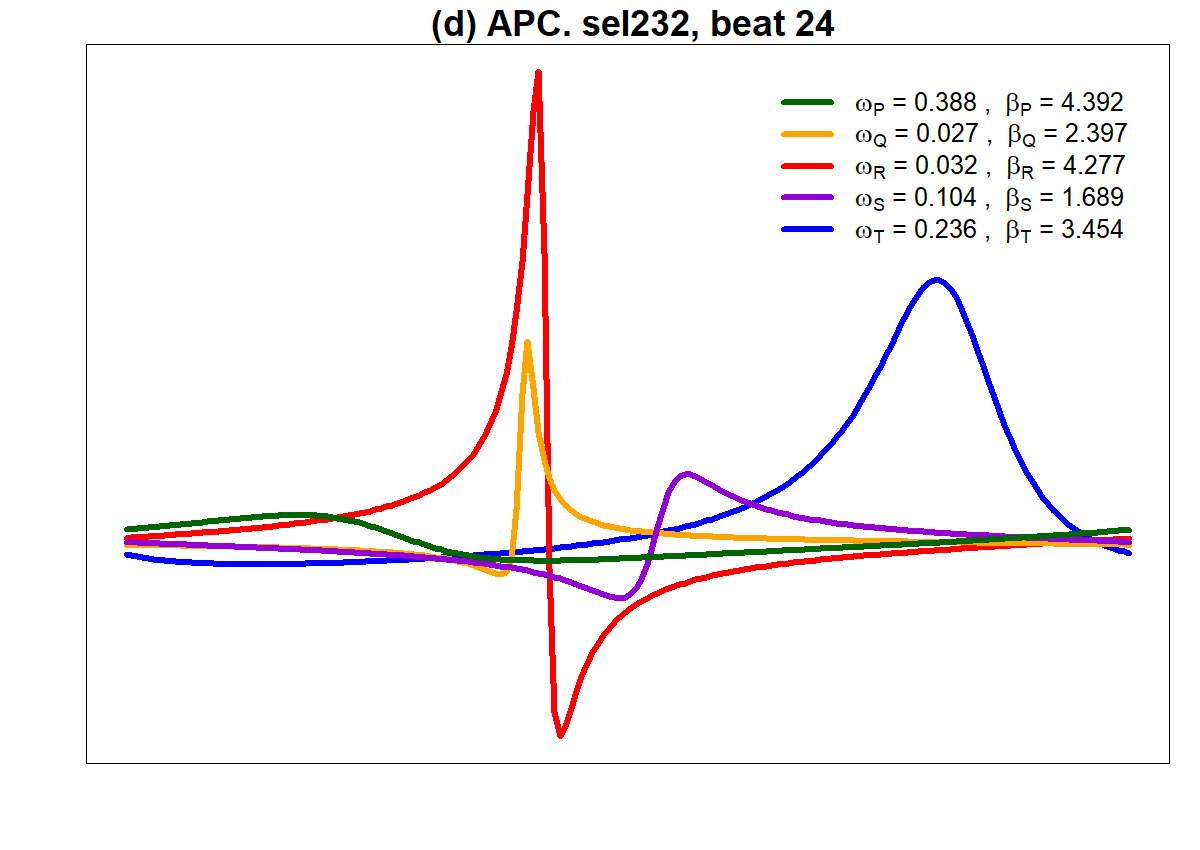}
  \label{f:Figure2sd}
\end{subfigure}\\
\begin{subfigure}{.5\linewidth}  
  \centering
  \includegraphics[width=1\textwidth]{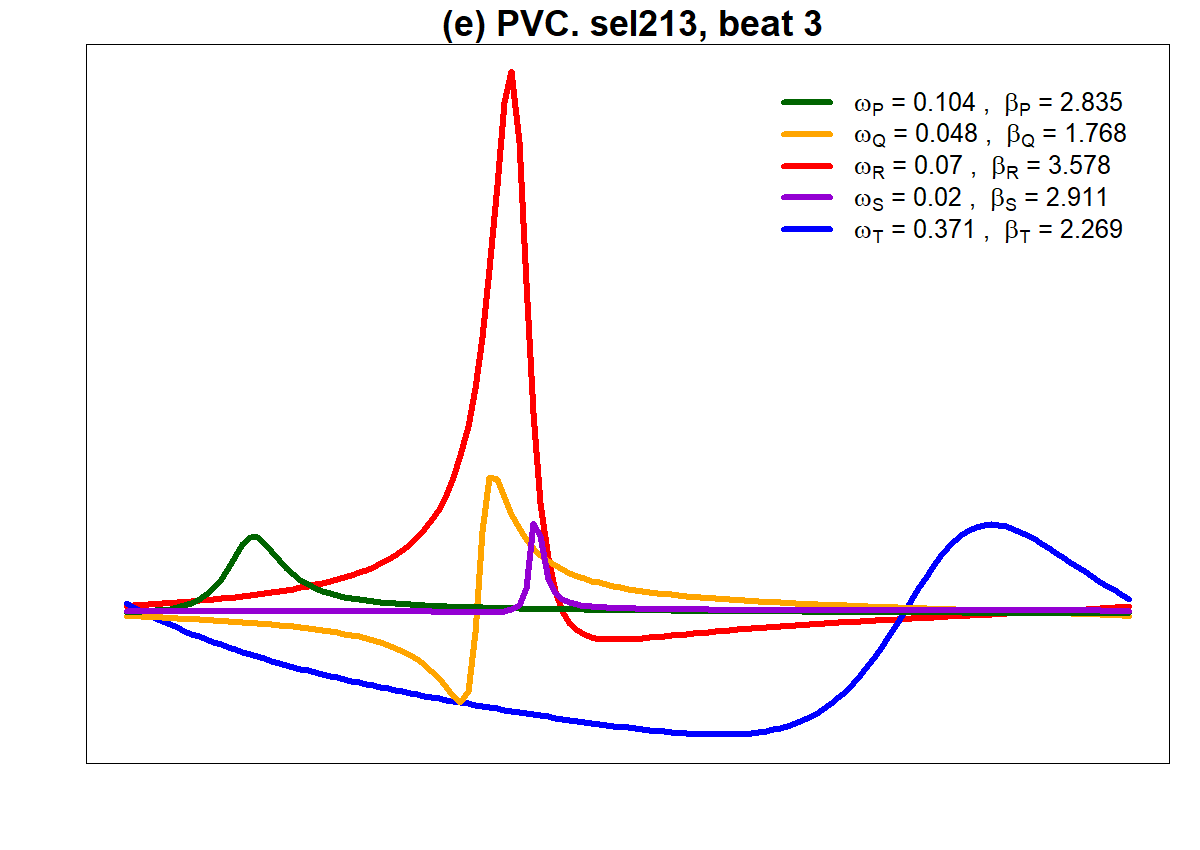}
  \label{f:Figure2se}
\end{subfigure}
\begin{subfigure}{.5\linewidth}  
  \centering
  \includegraphics[width=1\textwidth]{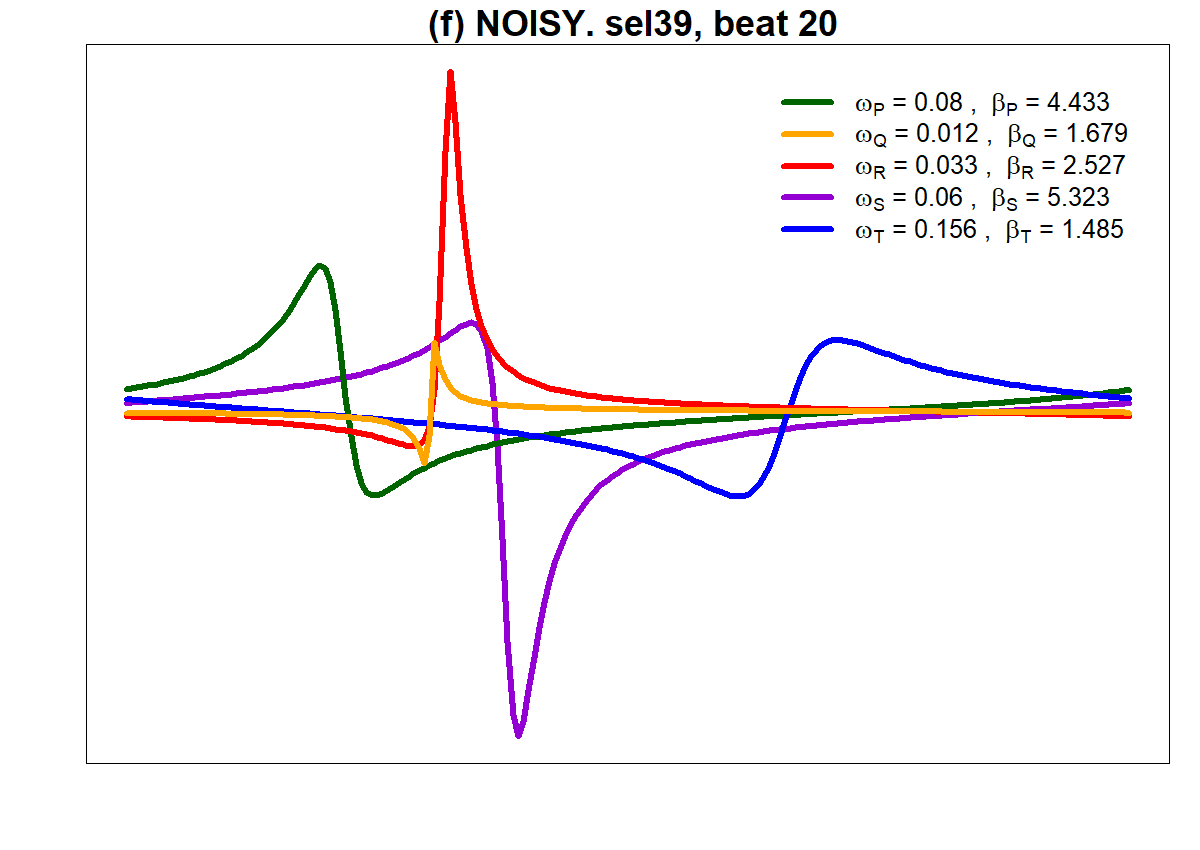}
  \label{f:Figure2sf}
\end{subfigure}%
\caption{$FMM_{ecg}$ waves and corresponding parameters, for representative beats from (a) NORMAL, (b) PACE, (c) RBBB, (d)APC, (e) PVC and (f) NOISY patterns. $P$ (green), $Q$ (yelow), $R$ (red), $S$ (violet), and  $T$ (blue)}
\label{f:Figure2s}
\end{figure}

It is interesting to observe how the specific shapes of the five main waves contribute to draw the observed pattern of the different morphologies as it is shown in  Figure \ref{f:Figure2s}. The estimated values of the parameters, recorded on the right side of the plots, quantify and describe the patterns, and explain the differences between the morphologies.
\\

On the other hand, the potential of the $FMM_{ecg}$ parameters to solve the problem of subject identification is also shown. A Fisher linear discriminant analysis is applied, using as predictors: $ A_J, \omega_J, \beta_j; J=P,Q,R,S,T$ (where missing values are replaced with the median value of the corresponding patient) and the one-leave-out rule to estimate the error rate. Only 8.6\% out of the 3,623 beats do not correctly identify the true patient. This error rate is very low taking into account the difficult task of discrimination among the 105 patients. As far as we know, this is the first time that this milestone has been achieved for the QT database, since other authors consider specifically selected sets of patients of a much smaller size (\cite{Sri19} and references therein). Moreover, a complete analysis is provided in the Supplementary Information, including specific-patient plots and statistics for the main $FMM_{ecg}$ parameters, see Figures S4-S10 and Table S4, respectively.  The results reveal consistency and reliability  of estimators and a great potential for individual identification tasks.

\subsection{Analysis of QT Data base signals. $P$ and $T$ wave annotations.}

This question is still a challenge as \cite{Elg15}, \cite{Yoc16}, \cite{Raj18} or \cite{Bho19},  among others, confirm.

Let $\widehat{t}_J^{FI}$, $J=T,P$ be the fiducial $FMM_{ecg}$ marks. Where if wave $J$ is positive ($\widehat{t}_J^{FI}=\hat{t}_J^{U}$) or negative ($\widehat{t}_J^{FI}=\hat{t}_J^{L}$) is determined by  $\widehat{\beta}_J$ and $\hat{\mu}({t}^{FI}_J).$

In order to perform a  fair comparison with alternatives approaches, we follow the analysis in \cite{Fri18}. Several measures are calculated to assess the wave detection: sensivity ($Se=\frac{TP}{TP+FN}$),  positive predictive value ($PPV=\frac{TP}{TP+FP}$), detection error rate ($DER=\frac{FP+FN}{TP+FN}$)  and F1 score ($F1=\frac{2TP}{2TP+FP+FN}$), where TP is the number of true positive detections, FN stands for the number of negative detections and FP stands for the number of false positive detections, that is, when the fiducial mark is outside the range of $\pm$75ms from the annotated peak/trough. 

Table \ref{t:PTwaves} shows the results, along with the four best methods in \cite{Fri18}, i.e., Martinez PT, Martinez WT+templates, Martinez WT+PT and Martinez PT + templates.

 \begin{table}[htbp]
\caption{ Summary of performance measures $P$ and $T$ waves detection from QT first signal data.}
\label{t:PTwaves}%
\resizebox{16cm}{!} {
\begin{tabular}{cccccccccc}
\toprule
Method & &\multicolumn{7}{c}{$P$ Wave} \\ \midrule
& No. beats&\textit{TP}&\textit{FP}&\textit{FN}&\textit{Se}(\%)&\textit{PPV}(\%)  &\textit{DER}(\%)  &\textit{F1}(\%)\\ \midrule
\textbf{$FMM_{ecg}$} &3194&3085&212&109&\textbf{96.59}&\textbf{93.57}&\textbf{10.05}&\textbf{95.05}\\
Martinez PT&3194&2859&342&335&89.51&89.32&21.20&89.41\\
 Martinez WT+templates&3194&2751&395&443&86.13&87.44&26.23&86.78\\
  Martinez WT+PT&3194&2932&416&262&91.80&87.57&21.23&89.64\\
Martinez PT+templates&3194&2816&320&378&88.15&89.90&21.85&89.41\\ \midrule
Method & &\multicolumn{7}{c}{$T$ Wave}\\ \midrule
 & No. beats&\textit{TP}&\textit{FP}&\textit{FN}&\textit{Se}(\%)&\textit{PPV}(\%)  &\textit{DER}(\%) &\textit{F1}(\%)\\\midrule
\textbf{$FMM_{ecg}$ } &3542&3542&415&0&\textbf{100}&\textbf{89.51}&\textbf{11.72}&\textbf{94.47}\\
Martinez PT &3542&2985&559&557&84.27&84.22&31.50&84.25\\
 Martinez WT+templates &3542&3115&464&427&87.94&87.03&25.15&87.49\\
   Martinez WT+PT&3542&3030&558&512&85.54&84.44&30.20&84.99\\
Martinez PT+templates&3542&3035&505&507&85.68&85.73&28.57&85.71\\
\hline
\end{tabular}}

\end{table}

$FMM_{ecg}$ gives the best results for all the validation measures and for both $P$ and $T$  peak/trough detection. It is especially striking that DER is less than halved in comparison to other methods for both $T$ and $P$ wave detection. The accurate detection of waves provided by $FMM_{ecg}$ is more valuable as the algorithm has not been specifically designed for this task, as it also serves other purposes.
 \\
 
Specific patient measures are given in  Tables S5 and S6. Besides, Figures S11-S16 show cases where the $FMM_{ecg}$ annotation is correct but is annotated mistaken as \textit{FN} or \textit{FP}. In some of those cases, what happens is that Physionet annotation uses the information from the second signal or from a close beat. In other cases, what happens is the $FMM_{ecg}$ annotation is more reasonable than, or as least as reasonable as, the Physionet annotation, although different. These cases indicate that the good $FMM_{ecg}$ results from Table \ref{t:PTwaves} could even be improved.

\section{Discussion}

From the methodological point of view, two contributions are proposed in this paper that have never been described before in the literature. On the one hand, a regression model with  multiple oscillatory components, which is formulated in terms of angular variables that represents the periodic movement of the waves, and that incorporates restrictions among the parameters, is considered. And, on the other hand, an MI original algorithm of estimation is designed. These methodological contributions have been proved here to be very relevant for their application in the description of the cardiac rhythm, but the potential is higher as they will likely be able to solve problems in other fields.
\\

As for the contributions to the automatic diagnosis of cardiovascular diseases and other clinical uses, the highlight of our approach is that it provides a set of new parameters and features with high descriptive potential which provides a concise analytical description of the morphology of the five main waves; specifically, its high capacity in human recognition has been demonstrated. Moreover, it is also very reliable even in abnormal and poor quality ECGs, it does not use training data and it works independently of preprocessing, scale and frequency. 

The $FMM_{ecg}$ parameters can be very useful to generate an automatic diagnostic by imitating the recognition skills of human beings, because estimated values under a given condition can be compared with reference values. In addition, the influence of such factors as age, gender, physical condition, medication, anatomic or genetic differences can be taken into account.
In fact, actual automatic diagnosis proposals fail due to two main causes; firstly, because different and unreliable measurements are used; secondly, because different problems in origin generate partially similar morphologies and, conversely, a certain anomaly is not associated with a single pattern. 
Using personalized reference ranges avoids false positives in diagnosis and subscribes to the global trend towards personalized medicine.

Moreover, the new parameters can be used in experimental essays to test medical and preventive strategies, to study the evolution of the heart's functioning, or in biometric identification.
 \\

The limitations of the approach, which are also challenges and extensions for future research, are sketched out next.

Firstly, a catalogue of interesting patterns  together with their parametric characterization must be elaborated in collaboration with an expert.  This question is partially addressed here, but a much more precise and detailed study is needed. This task  should  be done  by the incorporation of  identification algorithms from other leads.

Secondly, there are a few patterns, such as the Atrial Flutter, that do not fit well into the five main wave paradigm, but for which it is possible to design a specific algorithm. The analysis of multiple leads would also facilitate the wave identification task and provide more accurate results.

Finally, the incorporation of covariates, the definition of multivariate models and dynamic models, are statistical extensions to be studied that have several  applications in the clinic. Specifically, the covariates would serve to assess the influence of medication or the effect of interventions and multivariate and dynamic models would serve to describe spatio-temporal behaviours and model relationships between biological processes.

\section{Methods}

The application of our method for the QT database analysis and simulations assumes that $QRS$ annotations are provided. The detection of  the $QRS$ complex is a highly researched problem and well solved;  interesting references on the subject are \cite{Pan85}, \cite{Man12}, \cite{Che17}, \cite{Liu18} and \cite{Dov19}, among others.   The  $QRS$ annotations and RR values (distances between consecutive $QRS$ annotations), provided by Physionet, are used to select the specific segment corresponding to a single beat in our  data analysis. For a given $QRS$ annotation, $t^{QRS}$, let $RR_-$ and $RR_+$ be the RR obtained from the previous and the next $QRS$ annotation, respectively. Then, the input for the analysis of a single beat are the observations, $X(t_i)$, where $t_i \in [t^{QRS}-40\%RR_-,t^{QRS}+60\%RR_+], i=1,...,n$,  which before entering the algorithm, pass a trend removal step to reduce the influence of the low frequency noise, if necessary.

The MI algorithm, described below, uses these input data to derive predicted values for the voltage and features. 
\subsection{MI Algorithm}

\tikzstyle{pre} = [trapezium, , trapezium left angle=110, minimum width=2.5cm, trapezium right angle=110,text centered, draw=black, fill=white]
\tikzstyle{pro} = [trapezium, , trapezium left angle=70, trapezium right angle=70, text centered, draw=black, fill=white]
\tikzstyle{M} = [rectangle, rounded corners,minimum width=2.5cm, minimum height=1cm, text centered, draw=black, fill=blue!30]
\tikzstyle{I} = [rectangle, rounded corners,minimum width=5.5cm, minimum height=0.5cm, text centered, draw=black, fill=orange!30]
\tikzstyle{Q} = [circle, minimum width=0.75cm, minimum height=0.75cm, text centered, draw=black, fill=white]
\tikzstyle{aux} = [diamond, minimum width=0.5cm, minimum height=0.5cm, text centered, draw=white, fill=white]
\tikzstyle{arrow} = [thick,->,>=stealth]

\tikzset{
  -|-/.style={
    to path={
      (\tikztostart) -| ($(\tikztostart)!#1!(\tikztotarget)$) |- (\tikztotarget)
      \tikztonodes
    }
  },
  -|-/.default=0.5,
  |-|/.style={
    to path={
      (\tikztostart) |- ($(\tikztostart)!#1!(\tikztotarget)$) -| (\tikztotarget)
      \tikztonodes
    }
  },
  |-|/.default=0.5,
}

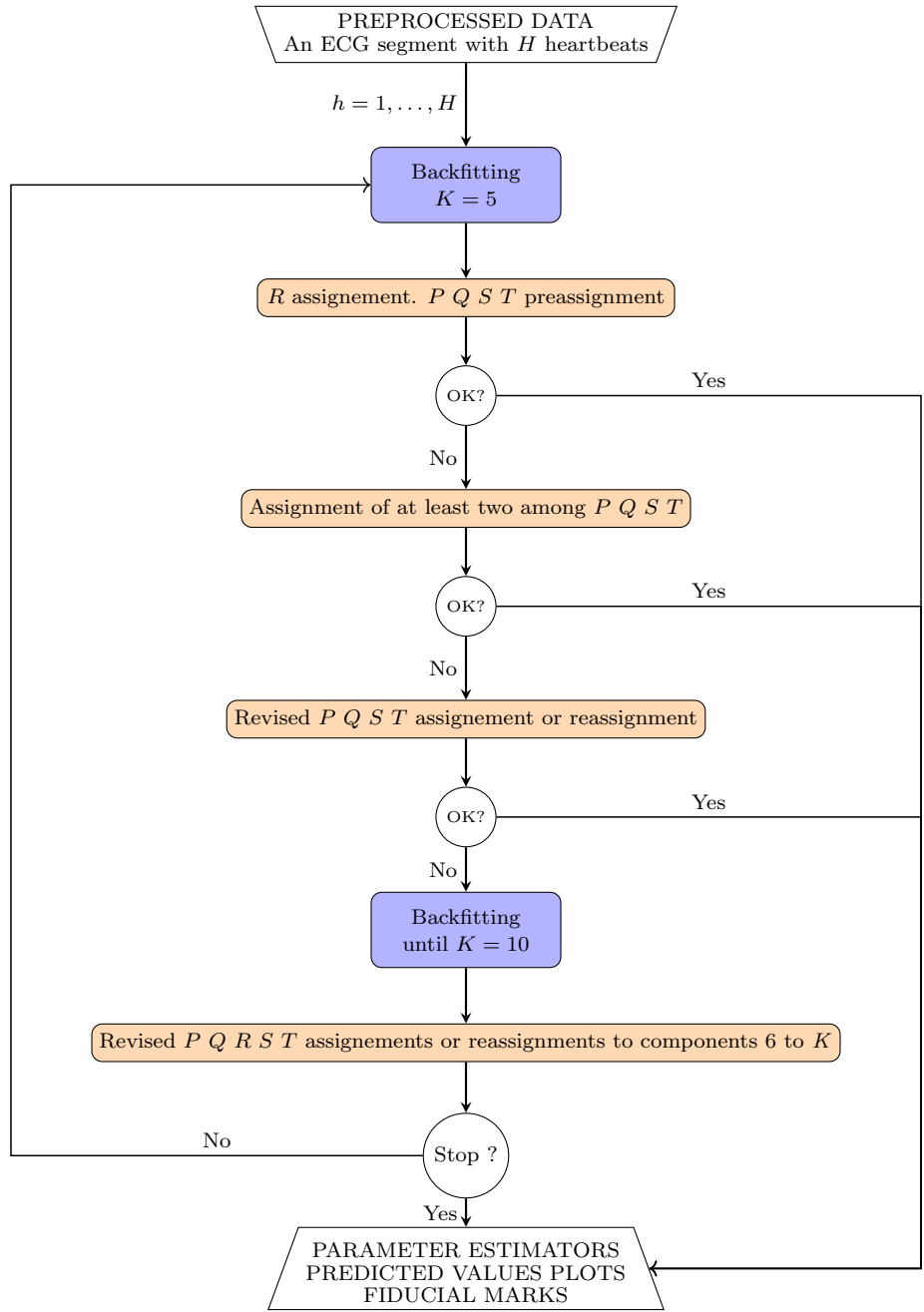
\begin{figure}[ht]
\begin{center}
\begin{tikzpicture}[node distance=1.5cm]

\scriptsize
\newcommand\arrowfromto[5][black]{%
  \draw[#1] #2 -- ( $ #2!#4!#3 $ ) node [midway, sloped, above] {#5}}
\node (input) [pre] {\shortstack{ PREPROCESSED DATA \\ An ECG segment with $H$ heartbeats }
	};
\node (M1) [M, below of=input, yshift=-0.5cm] {\shortstack{Backfitting\\$K=5$}};
\node (I1) [I, below of=M1]{\shortstack{$R$  assignement. $P$ $Q$ $S$ $T$  preassignment} };
\node (Q1) [Q, below of=I1, yshift=0.2cm] {\tiny{OK?}};
\node (I2) [I, below of=Q1] {Assignment of at least two among $P$ $Q$ $S$ $T$  };
\node (Q2) [Q, below of=I2, yshift=0.2cm] {\tiny{OK?}};
\node (I3) [I, below of=Q2] {Revised  $P$ $Q$ $S$ $T$ assignement or reassignment};
\node (Q3) [Q, below of=I3, yshift=0.2cm] {\tiny{OK?}};
\node (M2) [M, below of=Q3] {\shortstack{Backfitting\\ until $K=10$}};
\node (I4) [I, below of=M2] {Revised $P$ $Q$ $R$ $S$ $T$  assignements or reassignments to components 6 to \textit{K}};
\node (Q4) [Q, below of=I4] {\shortstack{Stop ? }};
\node (output) [pro, below of=Q4] {\shortstack{\\ PARAMETER ESTIMATORS\\PREDICTED VALUES PLOTS \\ FIDUCIAL MARKS}};
\node (aux1) [aux, left of=Q1, xshift=-2cm] {};
\node (aux2) [aux, left of=Q2, xshift=-2cm] {};
\node (aux3) [aux, left of=Q3, xshift=-2cm] {};
\node (aux4) [aux, left of=Q4, xshift=-2cm] {};
\draw [arrow] (input) -- node[anchor=east] {\scriptsize{$h=1,\dots,H$}} (M1);
\draw [arrow] (M1) -- (I1);
\draw [arrow] (I1) -- (Q1);
\draw [arrow] (Q1) -- node[anchor=east] {No} (I2);
\draw [arrow] (I2) -- (Q2);
\draw [arrow] (Q2) -- node[anchor=east] {No} (I3);
\draw [arrow] (I3) -- (Q3);
\draw [arrow] (Q3) -- node[anchor=east] {No} (M2);
\draw [arrow] (M2) -- (I4);
\draw [arrow] (I4) -- (Q4);
\draw [arrow] (Q4) -- node[anchor=east] {Yes} (output);
\draw[line width=0.2mm,->] (Q1) -- node[anchor=south] {Yes} ++(6 cm,0) |- (output);
\draw[line width=0.2mm,->] (Q2) -- node[anchor=south] {Yes} ++(6 cm,0) |- (output);
\draw[line width=0.2mm,->] (Q3) -- node[anchor=south] {Yes} ++(6 cm,0) |- (output);
\draw[line width=0.2mm,->] (Q4) -- node[anchor=south] {No} ++(-6cm,0) |- (M1);

\end{tikzpicture}

\end{center}
\caption{$FMM_{ecg}$ MI algorithm}
\label{f:alg}
\end{figure}

Consider the model in Definition \ref{eq:Mobecg}. The estimation problem reduces to solving the following optimization problem:
\begin{equation*}\label{eq:opt}
Min_{ \theta \in \Theta}  \sum_{i=1}^n [ X(t_i)-\mu(t_i,\theta)]^2
\end{equation*}

Where $\Theta$ is the parametric space. For a typical ECG pattern $\Theta$ is simply defined  as in Definition \ref{eq:Mobecg} through the restrictions among the $\alpha$'s.  However,  in order to arrive to a right identification of letters in atypical patterns in real practice, additional restrictions are needed. Mathematically, it means that $\Theta$ is reduced and are incorporated as thresholds in the algorithm.

The optimization problem above is computationally intensive and it is solved using a iterative algorithm which alternates  M and I  steps that provide successive estimators for  $W_J,J=P,Q,R,S,T$. The M step provides $K \ge 5$ oscillatory components using a backfitting algorithm  and  the I step assigns $K \le 5$ letters to, at most, five of these components. Typically, $K=5$,  however, in the presence of significant noise or when the morphology is pathological, sometimes, the interesting waves may be null or  be hidden between the sixth or seventh component (very exceptionally in others). For each component, the $FMM$ parameter values and percentage of explained variance, $\textbf{PV}$, are computed. The latter defined as follows,
\begin{equation*}\label{eq:PV}
PV_k=  R^2_{1,...,k}- R^2_{1,...,k-1},
\end{equation*}
where  $R^2_{1,...,k}$, defined in (\ref{eq:R}), refers to a multicomponent $FMM$ model with $K=k$ components.
For atypical patterns, the identification is done using thresholds which have been checked over many previous fits to a wide variety of ECG patterns in Physionet.

The initial values for the components to start the backfitting are those of the waves assigned so far and zero for the rest. 
The algorithm finishes when there is no significant increase in the percentage of variance explained or when a maximum number of iterations is attained. An increase of less than 0.01\% in the percentage of variance explained and a maximum of 10 iterations has been used in the analysis of the QT database.
\\

\textbf{M step}: The backfitting algorithm is designed by fitting a single $FMM$ component succesively to the residuals. To fit a single component, an adapted algorithm from that in \cite{Rue19} is developed.  The numbers of backfitting passes depends on the initialization. In the first M step up to 5 full turns of the backfitting are made.
\\

\textbf{I  step}: The $R$ is assigned in the first place. $R$ wave  corresponds to the component, in the top five, with highest \textit{PV} between components close to $t^{QRS}$, $\pi/2<\beta<5\pi/3$ (with a crest not a trough), $\omega<0.12$ (sharp) and maximum $\mu(t_J^U)$ (exceptionally the second maximum).
Next, preassignation of $P,Q,S$ and $T$ to the free components among the first five is done using $\alpha_P \le\alpha_Q \le \alpha_R\le \alpha_S \le \alpha_T$.  This preassigment corresponds to the  definite assignment in typical patterns. Successive steps are needed when the preaasignation components do not exhibit the expected wave morphology features, known from literature; it can be due to the absence of a wave or to the presence of noisy components. New assignations of letter to components are conducted using thresholds on the $FMM$ parameters that represent the previous knowledge. For instance, thresholds  to decide between $P$ or $Q$, are derived assuming that $Q$ is between $P$ and $R$ ($\alpha_P\le \alpha_Q \le\alpha_R$),  $Q$ is often sharper ($\omega_Q< \omega_P$), and $Q$ has a trough, while $P$ has a crest. Noisy components are detected with small \textit{PV}'s and $\omega$ values.

The outputs will be considered satisfactory (OK) only when the five letters are assigned and the parameters of the corresponding components describe the expected morphology.
\\

Figure \ref{f:alg} shows a flowchart of the algorithm where different colours are used for M and I steps.  The R code to implement the algorithm is available from corresponding author on reasonable request.

\section*{Acknowledgments}
The authors gratefully acknowledge the financial support received by
the Spanish Ministerio de Ciencia e Innovaci\'on and European
Regional Development Fund; Ministerio de Econom\'ia y Competitividad
grant [MTM2015-71217-R to CR] and Spanish Ministerio de
Educaci\'on, Cultura y Deporte [FPU14/04534 to YL].

\subsection*{Author contributions}
CR:  Conceived aims, theoretical proposal, conceptual design, designed of MI algorithm, developed computational code, wrote the manuscript.
YL : Designed of MI algorithm, developed computational code, processed original data, generated simulations, approved manuscript.
AL: Optimized the computational code, approved manuscript.

The authors declare no potential conflict of interests.

\section*{Materials \& Correspondence}

Materials and correspondence should be sent to	Dr. Rueda. email: \url{cristina.rueda@uva.es}

 \bibliography{referenciasECG}

\end{document}